# Emergence of quantum transport in SnTe films upon structural refinement

Shrinkhala Sharma[1*], Rabindra Basnet[1], Thomas Parker[2], Afolabi Akinfaderin[3], Alexander Samokhvalov[3], Rajeswari Kolagani[4], and Ramesh C. Budhani[1*]

[1]*Department of Physics, Morgan State University, Baltimore, Maryland 21251, USA.*
[2]*U.S. Army Research Lab, Aberdeen Proving Ground, Maryland 21005, USA.*
[3]*Department of Chemistry, Morgan State University, Baltimore, Maryland 21251, USA.*
[4]*Department of Physics, Towson University, Towson, Maryland, 21252, USA.*

Email: shrinkhala.sharma@morgan.edu and ramesh.budhani@morgan.edu

Topological crystalline insulators are distinguished from the conventional Z2 topological insulators by crystalline symmetry protection of the surface states. Quantitative characterization of these states requires magneto-transport measurements from which the phase coherence length and number of independent coherent channels can be extracted. Here we report such measurements on textured thin films of SnTe grown on $Bi_2Te_3$ buffered c-plane sapphire ($Al_2O_3$) by pulsed laser ablation. A post deposition annealing protocol was implemented to control the crystallographic texture, which reveals improved crystallinity and enhanced epitaxial alignment along the [111] direction in films annealed at elevated temperatures. The higher carrier density and lower resistivity seen in films annealed at 100 °C are suggestive of Sn vacancies which act as native acceptor defects, while films annealed at 200 °C approach the 1:1 stoichiometry. The magnetoresistance of all samples at T < 10 K characterized by a cusp centered around zero-field is suggestive of weak antilocalization and interpreted in the framework of the Hikami-Larkin-Nagaoka theory in the strong spin-orbit coupling limit. At T > 20 K, the MR is small and positive, and deviates from conventional Kohler scaling. An approximate collapse of the MR curves between 50 and 250 K is obtained using an empirical temperature-dependent scaling factor, indicating that changes in the Hall-derived carrier density alone cannot account for the observed scaling behavior. These results demonstrate that post growth annealing strongly influences the interplay between stoichiometry, crystallinity, and coherent transport in SnTe thin films.

## I. INTRODUCTION

Topological insulators (TIs) represent a distinct quantum state of matter characterized by an insulating bulk and metallic surface states protected by time-reversal symmetry [1] These topologically protected surface states exhibit spin-momentum locking and are robust against nonmagnetic disorder, making them attractive for applications in spintronics and quantum information technologies [2–4] . A related class of materials, known as topological crystalline insulators (TCIs), derives its topological protection from the crystal's mirror symmetry rather than time-reversal symmetry [4,5]. Among the theoretically predicted and experimentally confirmed TCIs, SnTe has emerged as the prototypical material because of its inverted band structure and mirror-symmetry-protected Dirac surface states. Angle-resolved photoemission spectroscopy (ARPES) measurements have confirmed these Dirac-like surface states on the (001) and (111) surfaces of SnTe, establishing it as a model system for investigating topological crystalline phases [6,7].

Low-temperature magnetotransport measurements provide a powerful approach for probing topological surface states (TSS). A key feature of transport in systems with TSS is weak antilocalization (WAL), a quantum interference effect associated with a Berry phase of π. Owing to strong spin-orbit coupling, the electron spin is locked to its momentum and rotates as the electron traverses a closed path, causing its wave function to acquire this additional phase. In conventional conductors, constructive interference between time-reversed electron trajectories enhances backscattering, resulting in weak localization (WL) and a reduction in electrical conductivity [8]. In contrast, the π Berry phase associated with spin-momentum-locked surface states causes these trajectories to interfere destructively, thereby suppressing backscattering and enhancing conductivity near zero magnetic field [9]. An applied magnetic field breaks time reversal symmetry and introduces an additional phase difference between the time-reversed trajectories, weakening their destructive interference and suppressing the WAL correction. Analysis of the magnetoconductance using the Hikami–Larkin–Nagaoka (HLN) model [10] enables the extraction of key transport parameters, including the phase coherence length and the number of coherent conduction channels, providing valuable insights into the interplay between surface and bulk transport.

In addition to its TCI nature and attendant quantum transport, SnTe crystals also exhibit unique ferroelectric (FE) properties on transitioning from the high temperature rock salt structure to a low temperature rhombohedral phase on cooling [11]. In ultrathin films, robust in-plane ferroelectricity can persist down to the monolayer limit. Tin telluride has also been recognized as a strong spin-to-charge conversion material [12,13], making it a potential candidate for spintronic applications and its high photo-thermoelectric response from visible to infrared spectral range make thin films of SnTe suitable for detector applications [14,15] .

SnTe thin films have been synthesized using a variety of techniques, including molecular beam epitaxy [16–20], vapor transport [21], and sputtering [22]. Among these, pulsed laser deposition (PLD) offers several practical advantages, such as precise control over film thickness, relatively low growth temperatures, operation at high background pressure of desired gases, and compatibility with large-area substrates. Although PLD has been widely employed for the growth of oxide [23–27] and chalcogenide thin films [28,29], comparatively fewer studies have investigated the low-temperature quantum transport properties of PLD-grown SnTe, particularly the influence of post-deposition annealing on its electronic transport behavior [30–32]. To date, we identified three relevant reports on the use of PLD for SnTe-based thin films. Early work by Teghil et al [31]. investigated the laser ablation of SnSe and SnTe examining the plume composition and its implications for thin-film deposition. Oszwaldowski et al. described a PLD apparatus for semiconductor thin-film growth, including the preliminary deposition of SnTe thin film [32]. More recently, Si et al. [30] reported epitaxial indium - doped SnTe thin films grown by PLD on $BaF_2$ substrates, investigating their structural and superconducting properties. The influence of post-deposition annealing on the low – temperature transport properties of PLD-grown SnTe films remains largely unexplored.

In this work, we report the growth of epitaxial SnTe thin films on c-plane sapphire substrates using pulsed laser deposition, followed by *in-situ* post-deposition annealing at 100, 150, and 200 °C. The structural quality of the films is examined using X-ray diffraction and Raman spectroscopy, while their electrical transport properties are investigated through temperature-dependent resistivity, Hall effect, and magnetoresistance measurements. Particular emphasis is placed on understanding the evolution of weak antilocalization with annealing temperature through magnetoconductance analysis using the HLN model. The results demonstrate that thermal annealing significantly influences the crystalline quality and quantum transport properties of SnTe, providing insights into the relationship between structural order, phase-coherent transport, and the contribution of topological surface states.

## II. EXPERIMENTAL METHODS

SnTe thin films were grown on c-plane sapphire ($Al_2O_3$) substrates using a KrF excimer laser ($\lambda$ =248 nm) based PLD system. The substrates were cleaned in an ultrasonic bath of acetone and isopropanol and then annealed in a tube furnace at 1000 °C for 4 hours in the presence of oxygen. Prior to SnTe deposition, a 10-nm-thick $Bi_2Te_3$ buffer layer was deposited *in situ* onto the $Al_2O_3$ substrate by thermal evaporation at a rate of 0.4 Å/sec. This buffer layer was introduced to reduce the lattice mismatch between the substrate and SnTe from approximately 6% to 2%, thereby promoting improved structural quality and epitaxial growth. For SnTe deposition, a 100 % dense, melt-grown target with a nominal composition of $SnTe_{2.2}$ and an elemental purity of 99.999% was used. The target was ablated at a repetition rate of 3 Hz with an areal laser energy density per pulse of ~ 2 J/cm$^2$ yielding a growth rate of ≈ 0.5 Å/s in 2 mTorr background pressure of ultrahigh purity argon. During deposition, the substrate was continuously rotated at 4 Hz to ensure uniform film growth, while the target was rotated at 20 Hz in addition to rastering the laser spot to minimize localized ablation and maximize the usage of the target. The films were deposited at room temperature with a nominal thickness of 50 nm and thereafter annealed at 100 °C, 150 °C and 200 °C under an argon atmosphere. Significant film loss was observed upon annealing at 250 °C, while the SnTe film completely evaporated from the substrate surface at 300 °C. The crystal structure and phase of the thin films were characterized using a Rigaku SmartLab high – resolution X-ray diffractometer (XRD). Surface morphology and elemental composition were investigated using a JEOL JCM -7000 scanning electron microscope (SEM) operating at an accelerating voltage of 15 kV and a working distance of 13.8 mm under high vacuum, equipped with an energy-dispersive

X-ray spectroscopy (EDS) system. X-ray photoelectron spectroscopy (XPS) measurements were performed using a PHI Gemini XPS system using a monochromatic Al Kα radiation ($h\nu$ = 1486.6 eV), an irradiated spot size of 100 x100 $\mu m^2$, and an X- ray power of 25 W. The take-off angle relative to the surface normal was 54.7°. High resolution scans were conducted with a pass energy of 27 eV and a step size of 0.1 eV. The survey scans were conducted with a pass energy of 244 eV and a step size of 1 eV. These measurements were carried out at a base pressure of 7.5 x $10^{-10}$ Torr. Raman spectroscopy measurements were performed using a Horiba Confocal Raman Microscope with a 532-nm excitation laser and a 50X objective. Electronic transport measurements were carried out in a physical property measurement system (PPMS; Quantum Design) equipped with a 9-T superconducting magnet. Samples for these measurements were patterned into Hall-bar geometries.

## III. RESULTS AND DISCUSSION

While the isostructural PbTe does not host a topologically non-trivial phase, SnTe, does. The {001}, {110} and {111} surfaces of face centered cubic SnTe have mirror symmetry with respect to the (110) mirror plane [33]. Unlike conventional Z2 topological insulators, where band inversion occurs at a single time reversal invariant point, SnTe's band inversion occurs simultaneously at four symmetry equivalent L points of its Brillouin zone [6,7].

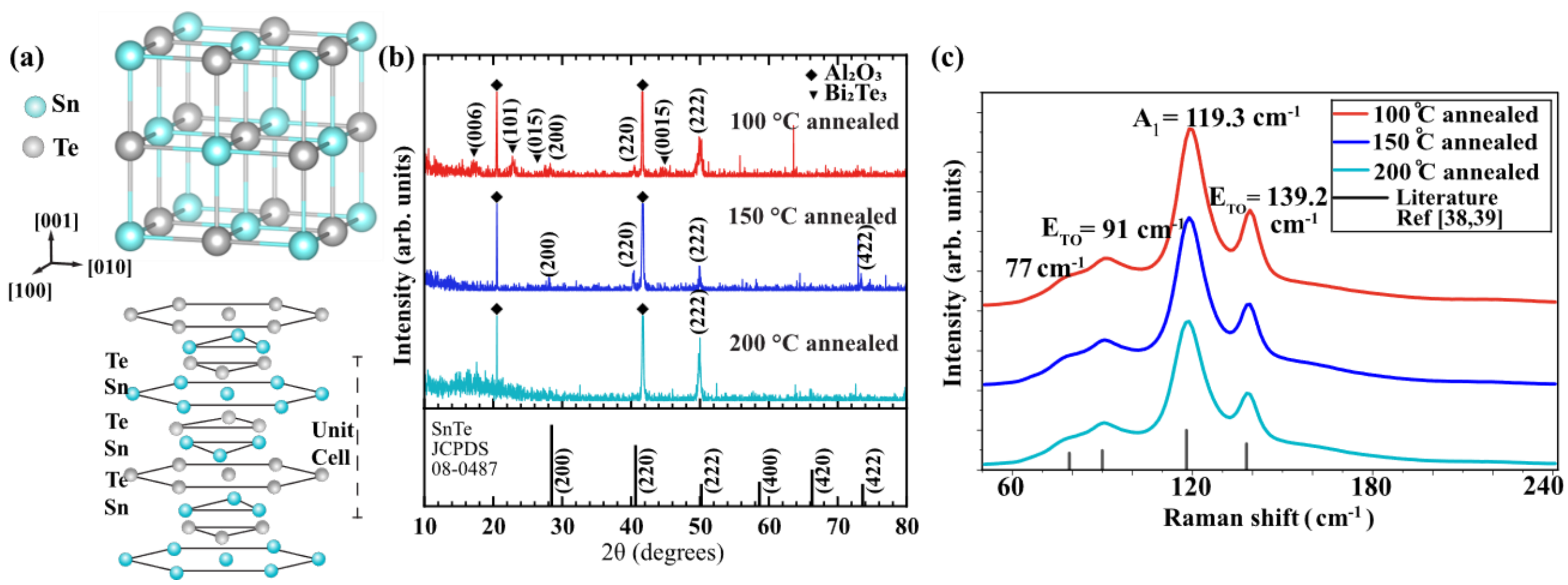

FIG.1. Structural and vibrational characterization of SnTe thin films. (a) Crystal structure of rock salt SnTe with alternating Sn and Te atoms occupying two interpenetrating face centered cubic lattices (top), and the alternating stacking of Sn and Te layers along the [111] direction (bottom). (b) X- ray diffraction patterns of the SnTe thin films annealed at 100, 150 and 200 °C, confirming the formation of SnTe phase. (c) Room temperature Raman spectra of the corresponding SnTe films, showing the characteristic vibrational modes of the SnTe cubic phase. The vertical black lines mark the mode positions reported in the literature [38,39].

Under equilibrium conditions, SnTe crystallizes in the face-centered cubic (FCC) phase with the space group $Fm\overline{3}m$ (No. 225) with a lattice constant a = 0.631 nm. In this structure, Sn and Te atoms occupy alternating FCC sublattices, with each atom octahedrally coordinated by six atoms of the opposite species. Dirac surface states have been detected on both the (001) and (111) surfaces of SnTe using angle-resolved photoemission spectroscopy [7,34]. Notably, the surface state Dirac points behave differently depending on orientation: on the (001) surface they sit away from the time reversal invariant momenta (TRIM), whereas on the (111) surface all the four Dirac cones coincide with the TRIM points [35–37]. As a result, the (111) surface states carry an additional layer of protection, arising from the time reversal symmetry on top of the underlying crystal symmetry [35] . Viewed along the [111] direction, this structure can equivalently be described in a hexagonal setting as a stack of alternating Sn and Te layers, as shown in the bottom panel of Fig. 1(a). The omega–two theta (ω–2θ) XRD scans of the films presented in Fig. 1(b) can be indexed to the cubic SnTe phase, confirming the successful growth of phase-pure SnTe without detectable secondary phases. A clear evolution of the diffraction pattern is observed with increasing annealing temperature. The films annealed at lower temperatures exhibit multiple diffraction peaks corresponding to different crystallographic orientations, whereas the intensity of the (111) reflection increases progressively with annealing temperature, indicating the development of a stronger (111) preferred orientation. Simultaneously, the diffraction peaks become sharper and more intense, suggesting improved crystallinity.

Figure 1 (c) shows the Raman spectra of the same films, providing vibrational fingerprints that complement the XRD analysis. In the ideal rock-salt SnTe structure, Sn and Te atoms sit on centers of inversion symmetry, so the zone-center optical phonons are formally Raman-inactive; the modes observed here become Raman active because of the local symmetry breaking around Te sites caused by the intrinsic Sn vacancies characteristic of p-type SnTe [38]. These optical modes involve opposite, out-of-phase Sn and Te displacements. The peaks at 91cm$^{-1}$ and 139 cm$^{-1}$ are attributed to the SnTe related transverse optical (TO) vibrations in which the relative motion of the Sn and Te atoms lies in the plane perpendicular to the wavevector [39,40] . The feature near 119 cm$^{-1}$ is associated with the $A_1$ symmetry longitudinal - optical (LO) mode, in which the atoms undergo relative displacement along the phonon wavevector [41] . The lower wavenumber feature at 77 cm$^{-1}$ may originate from defect-activated or vacancy- related vibrations. These results are consistent with the previously published works on SnTe films. We also performed the XPS measurements (Fig.S1of the Supplemental material [42] ) on our films. The peaks present at 485.5 eV and 493.8 eV correspond to the $3d_{5/2}$ and $3d_{3/2}$ spin orbit doublet of Sn. While, the Te $3d_{5/2}$ and $3d_{3/2}$ peaks appear at 571.44 eV and 581.89 eV respectively, consistent with reported peak positions of SnTe thin films [43,44]. The $3d_{5/2}$ binding energies, however are shifted from the elemental metals (Sn ~484.9 eV, Te ~ 573 eV) [45] toward values reported for SnTe single crystals (Sn $3d_{5/2}$ 485.6 eV, Te $3d_{5/2}$ 572.3 eV) [46] . This chemical shift indicates a well-formed SnTe compound rather than a mixture of Sn and Te. SEM-EDS measurements (Fig.S2 of the Supplemental Material [42]) show that the 200 ˚C and 150 °C annealed samples have a Sn:Te ratio close to 1:1, whereas the 100 °C film is Te-rich, with a Te:Sn ratio of approximately 2:1. These results show that the annealing temperature modifies the structural orientation, crystallinity and chemical composition of the SnTe films.

Following the structural analysis, we address magneto-transport in the SnTe films. Figure 2(a) shows the temperature-dependent resistivity ($\rho_{xx}(T)$) of the films annealed at 100 °C, 150 °C, and 200 °C. We find that all samples exhibit metallic behavior, with resistivity increasing monotonically with temperature, with the residual resistivity ratio, RRR = $\rho_{xx}$ (300 K)/$\rho_{xx}$ (T→ 0), of 1.4, 1.25 and 1.1 for the 100º, 150º and 200º C films, respectively. The Hall resistivity $\rho_{xy}$ of the film annealed at 200 °C shown in Fig. 2(d), is linear in field up to 9 T at all measured temperatures as can be seen more clearly in the magnified high-field region shown in the inset. The positive slope indicates hole-dominated transport, consistent with p-type doping induced by native Sn vacancies in SnTe [47] . The Hall data for the 150 °C and 100 °C film also remain linear in field (Figs. S3(b) and S3(c) of the Supplemental Material [42] ). We therefore analyze the data within an effective single-carrier model [19,48] . The slope of the $\rho_{xy}$ (B) gives the Hall coefficient $R_H$, from which we obtain the carrier concentration $n_H = 1/(e|R_H|)$. Combining $R_H$ with the zero-field resistivity gives the Hall mobility $\mu_H = |R_H|/\rho_{xx}$. Figures 2(e) and 2(f) show that the 100 °C film has the highest carrier concentration and mobility. Both quantities decrease with increasing annealing temperature, consistent with the corresponding increase in resistivity within the single-carrier picture. The Hall carrier concentrations remain on the order of $10^{20}$ $cm^{-3}$ across the measured temperature range. The carrier density is largest and most weakly temperature dependent in the 100 °C film (~5.5x$10^{20}$ $cm^{-3}$ at low temperature, decreasing to ~5x$10^{20}$ $cm^{-3}$ at 300 K), while the 150 °C and 200 °C films show systematically lower and more strongly decreasing carrier densities (~2.5-3.6 x $10^{20}$ $cm^{-3}$). These values fall within the range reported for SnTe films irrespective of the growth technique: from ~2x$10^{17}$-1x$10^{21}$$cm^{-3}$ in MBE films [18,19,49] , and ~5.5x$10^{20}$ $cm^{-3}$ - 3.4x$10^{21}$ $cm^{-3}$ in annealed sputtered polycrystalline films [22], indicating that the carrier concentration in our PLD films is in good agreement with the existing literature. The mobility does not scale inversely with anneal temperature in the same manner: the 100 °C film reaches ~31 $cm^2V^{-1}sec^{-1}$ at low temperature, decreasing to ~23 $cm^2$ $V^{-1}sec^{-1}$ at 300 K, whereas the 150 °C and 200 °C films remain substantially lower and nearly temperature independent (~15-17 $cm^2V^{-1}sec^{-1}$). These values are lower relative to those of epitaxial films which range from ~ 60-720 $cm^2V^{-1}sec^{-1}$ [19] depending on the growth temperature, but are comparable to, or somewhat higher than, those reported for annealed sputtered SnTe films (~7-13 $cm^2V^{-1}sec^{-1}$) [22]. Although PLD-grown SnTe films have also been reported, temperature-dependent Hall carrier densities and mobilities are not as extensively documented, limiting a direct comparison with the present results [30,31] .

To highlight the low – temperature behavior of the electrical resistivity, we present in Fig. 2(b) an enlarged view of the $\rho_{xx}$ (T) of the 200 °C annealed film. A minimum in resistivity is seen near 125 K, followed by a second shallow dip centered around ~8 K upon further cooling. Since the carrier concentration and Hall mobility both evolve smoothly over the same temperature range, these anomalies cannot be explained solely within the classical Drude framework, in which the resistivity is determined by the product of carrier concentration and mobility. Instead, the low-temperature upturn suggests the emergence of a magnetic-field-sensitive correction to the conductivity. To distinguish the origins of the two resistivity minima, we first examine the low-temperature feature near 8 K by measuring the longitudinal resistivity from 2.5 to 60 K under an out-of-plane magnetic field. These data, collected at fields ranging from 0.03 T to 9 T are shown

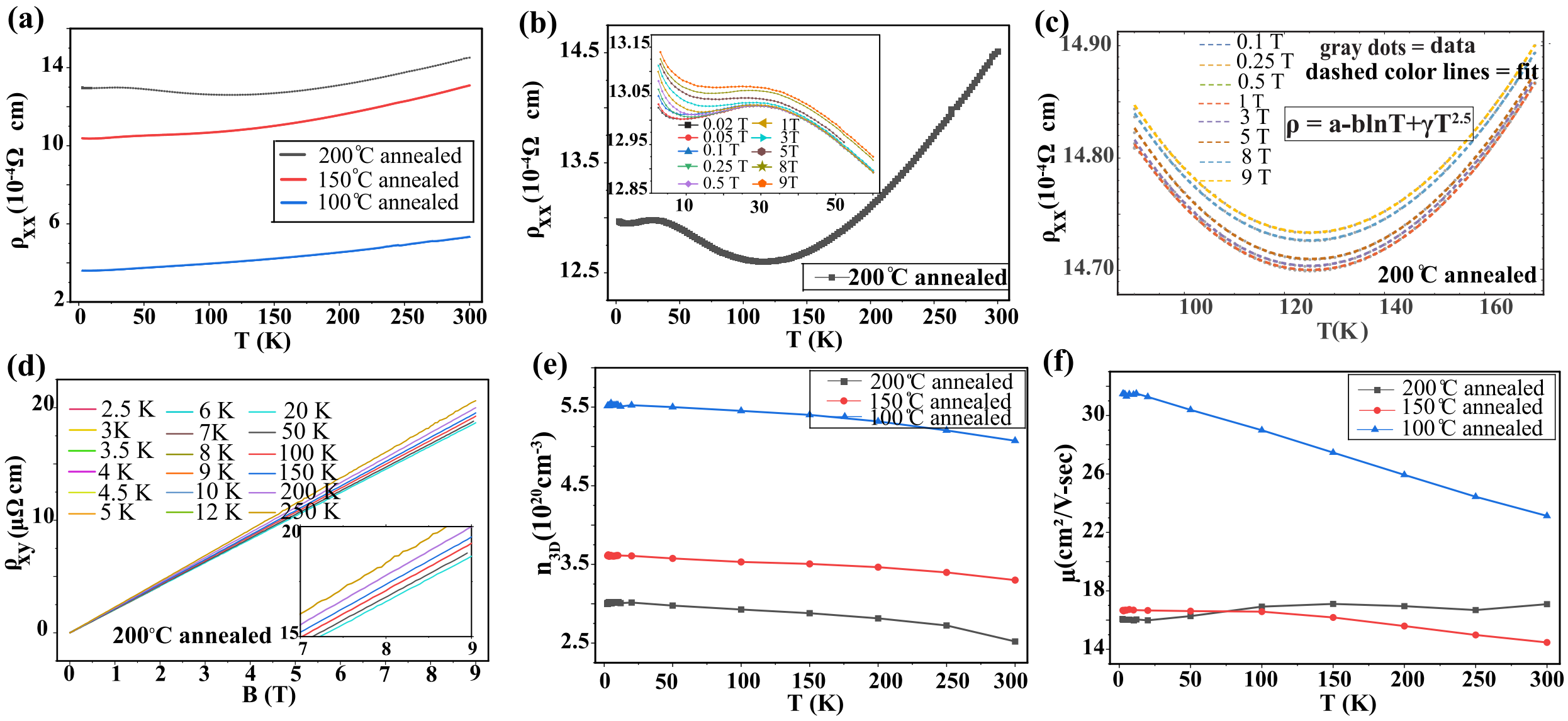


FIG.2. Temperature dependent electrical transport properties of PLD grown SnTe thin films annealed at 100°, 150° and 200° C. (a) Longitudinal resistivity, $\rho_{xx}$, as a function of temperature (2.5- 300 K) for the three annealing conditions. (b) Expanded view of the low temperature resistivity of the 200°C annealed film, showing a shallow resistivity minimum followed by a weak upturn upon further cooling. Inset: Temperature dependence of $\rho_{xx}$ measured under out of plane magnetic field from 0.02 to 9T. (c) expanded view of the broader resistivity minimum in the 200 °C film under out of plane magnetic fields from 0.1 to 9 T. Gray dots are experimental data and dashed lines are fits to $\rho = a - b\ln T + \gamma T^{2.5}$. (d) Hall resistivity $\rho_{xy}$ as a function of magnetic field measured at selected temperatures for the 200°C annealed film. Inset: expanded view of the high-field region between 7 and 9 T. (e) Temperature dependence of the Hall carrier concentration, $n_{3D}$, extracted from the linear Hall response for all three annealing conditions. (f) Temperature dependence of the Hall mobility, μ, for the corresponding films.

in the inset of Fig. 2(b). The sample shows a marked positive magnetoresistance and the minimum at ~ 8 K becomes shallower with increasing magnetic field. This positive out-of-plane MR is suggestive of a quantum correction to conductivity due to weak antilocalization effects in the sample, which will be discussed later in the paper. The positive MR decreases with increasing temperature, from 0.9 % at 2 K to 0.3 % at 60 K under an applied field of 9 T, consistent with the suppression of quantum interference effects at elevated temperatures. A similar upturn in resistivity at lower temperatures has been reported in SnTe quantum wells [50].

In contrast to the low-temperature feature near 8 K, the broader resistivity minimum near 125 K is only weakly affected by magnetic field, suggesting a distinctly different physical origin. To examine this feature more closely, Fig. 2 (c) shows an expanded view of $\rho_{xx}$ (T) between 90 and 170 K. On cooling from room temperature, the resistivity of the 200 °C film initially decreases, as temperature-dependent scattering is reduced; however, below $T_{min}$, this trend is reversed and a weak resistivity upturn develops. As discussed in Note S1 of the Supplemental Material [42] , this behavior is well described by $\rho_{xx} = a - b\ln T + \gamma T^p$ ,with p~2.5 reflecting competition between a metallic temperature dependent contribution and an interaction-induced logarithmic correction. The $\gamma T^p$ term describes the metallic contribution, which decreases upon cooling, whereas the logarithmic term can be understood within the metallic granular-transport framework.

In this regime, the conducting SnTe grains remain effectively coupled to support metallic transport, while electron transfer between neighboring grains is influenced by the finite charging energy associated with charge fluctuations on individual grains [51–53]. Coulomb interactions consequently renormalize the effective intergrain tunneling conductance, with the correction becoming progressively stronger upon cooling. This interaction effect produces a logarithmic reduction in conductivity with decreasing temperature [51–53]; when the correction is small compared with the background conductivity, it appears as an approximately -ln T contribution to the resistivity. The resistivity minimum therefore marks a crossover between the decreasing metallic contribution and the increasing influence of Coulomb interactions on intergrain transport. The nearly unchanged $T_{min}$ over the investigated magnetic field range further distinguishes this broad feature from the low-temperature anomaly. As shown in Fig. S3(a) of the Supplemental Material [42], neither the logarithmic nor the metallic contribution alone reproduces the minimum. An Efros-Shklovskii-type hopping description, characteristic of strongly localized granular transport [53,54], also does not describe the observed temperature dependence. A Reiss-type grain-boundary model [55,56], in which the temperature dependence arises primarily through the electron mean free path at fixed barrier transparency, does not give a unique or satisfactory account of the observed minimum. In contrast, the combined logarithmic and metallic form captures both sides of the minimum.

Next, we address the magnetic field dependence of $\rho_{xx}$ at fixed temperatures in the range of 2.5 to 10 K, where a distinct minimum followed by an upturn is seen upon cooling (inset Fig. 2(b)). A quantum coherent WAL as suggested by the low-field positive MR and highlighted by the cusp-like feature of magnetoconductance at 2.5 K, is apparent in Figs. 3(a)-3(c) for the 200, 150, and 100 °C annealed films respectively. In SnTe, WAL is often associated with spin-momentum-locked topological surface states, where destructive quantum interference suppresses backscattering and produces a conductance maximum near zero magnetic field [18,19]. However, a three-dimensional bulk contribution to WAL has also been reported in SnTe thin films [57] and In-doped SnTe nanoplates [58]. The data presented in Fig.3 are interpreted in the framework of the WAL theory developed by Hikami, Larkin and Nagaoka (HLN Model) [10,59] commonly applied to describe quantum corrections to conductivity from topological surface states [18,60] . The change in sheet conductance $\Delta G_s = G_s(B)-G_s(B=0)$, as a function of magnetic field B is given by the HLN model as,

$$\Delta G = -\frac{\alpha e^2}{\pi h}\left[\psi\left(\frac{\hbar}{4e\mathcal{L}_\phi^2 B}+\frac{1}{2}\right)-ln\left(\frac{\hbar}{4e\mathcal{L}_\phi^2 B}\right)\right]+\beta B^2 \qquad (1)$$

Here, $\psi$ is the digamma function, $\mathcal{L}_\phi$ is the phase coherence length which gives the average distance travelled by an electron between two successive inelastic scattering events, and α is a prefactor whose value reflects the number of independent coherent transport channels contributing to the WAL correction and depends on the nature of electron scattering. The topological surface states of SnTe belong to the symplectic class and each independent channel therefore contributes $\alpha = 0.5$ per independent channel [18,61], so that the number of channels is $N=2\alpha$. The coefficient

$\beta = \beta_q + \beta_c$ is a quadratic correction [59] . Its quantum part $\beta(\beta_q)$ captures both spin orbit coupling and elastic scattering in the HLN model and is given by,

$$\beta_q B^2 = -\frac{e^2}{24\pi h}\left[\frac{B}{B_{SO}+B_e}\right]^2 + \frac{3e^2}{48\pi h}\left[\frac{B}{(4/3)B_{SO}+B_\phi}\right]^2 \qquad (2)$$

and $\beta_c = \mu^2 B^2$ is the classical cyclotronic MR, where μ is the Hall mobility. The characteristic fields in Eq. (2), $B_e$, $B_\phi$ and $B_{SO}$ are linked to the elastic, phase coherence, and spin-orbit lengths through $B_n = \hbar/4e\, L_n^2$, where n = e, ϕ and so. The elastic mean-free-path $L_e$ was determined independently from the measured Hall carrier density and mobility for each film. The HLN equation with β = 0 is valid if the characteristic transport field $B_{tr}$ obeys the condition $B_{tr} >> B$ where $B_{tr} = \frac{\hbar}{2eL_e^2}$.

**Figure 3** shows the model fits for the films deposited at room temperature and annealed at the three temperatures. As expected, the WAL disappears at higher temperatures due to decrease in the

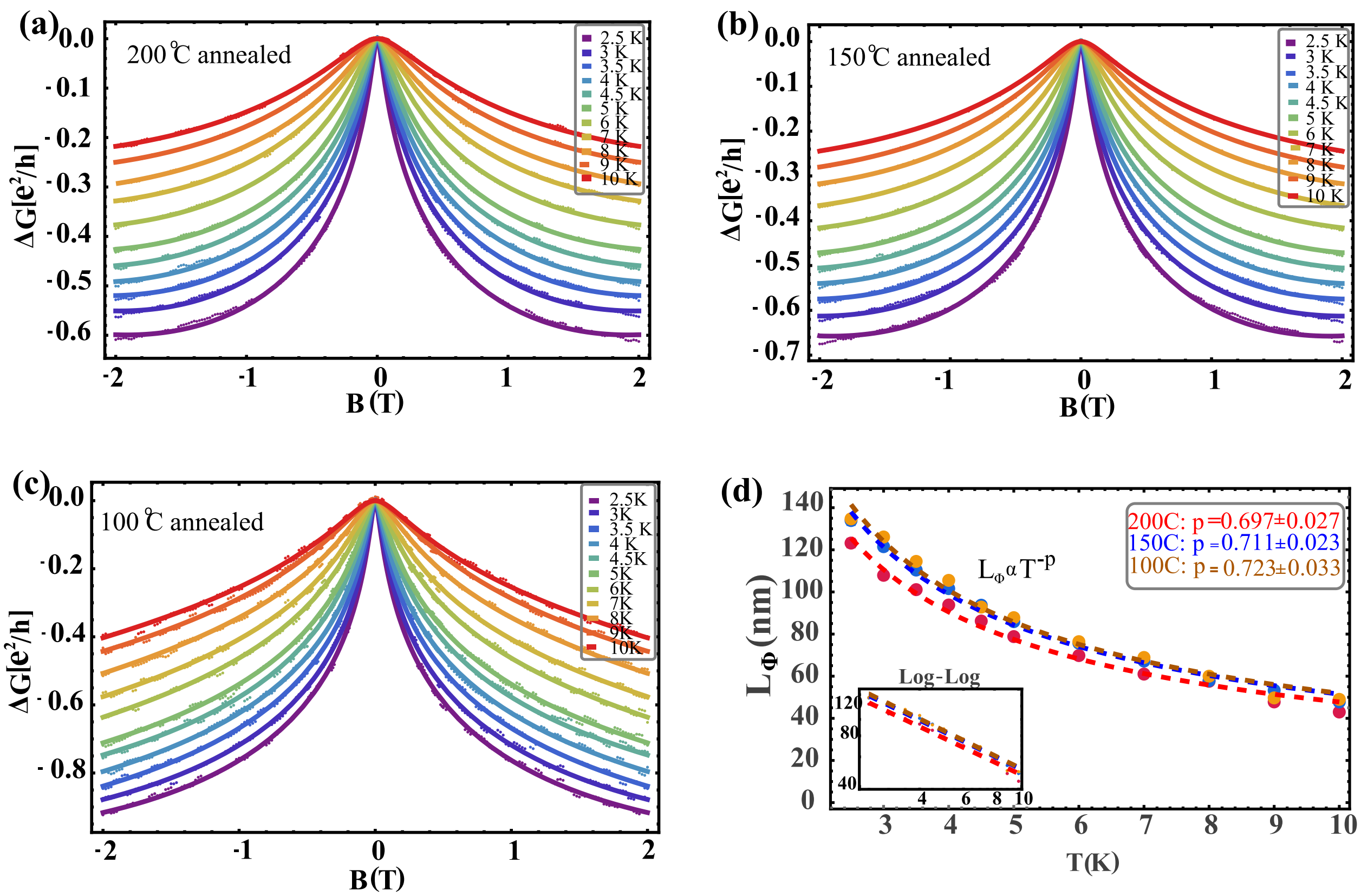


FIG.3. Weak antilocalization (WAL) in SnTe thin films annealed at three temperatures. (a-c) Change in sheet magnetoconductance ΔG (B) = G (B)-G (0), as a function of magnetic field at temperatures from 2.5 to 10 K, for the films annealed at 200, 150 and 100 °C. Solid curves are fits to the Hikami-Larkin-Nagaoka model, from which the phase coherence length $\mathcal{L}_\varphi$, the number of coherent channels 2α, and the quadratic coefficient β are extracted at each temperature. (d) Temperature dependence of $\mathcal{L}_\varphi$ for all the three annealing conditions, fit to the power law $\mathcal{L}_\varphi \propto T^{-p}$ (dashed curves), with extracted exponents p = 0.697±0.027, 0.711±0.023, and 0.723±0.033 for the 200, 150 and 100 C samples, respectively. Inset: same data and fits on a log-log scale.

phase coherence length caused by increased thermal scattering. This model allows us to extract the phase-coherence length $\mathcal{L}_{\phi}$, as well as the effective number of coherent transport channels 2α. From Fig. 3 (d) we see that the 100 °C annealed film exhibits the largest phase-coherence length, with $\mathcal{L}_{\phi} = 134\ nm$ at 2.5 K. The 200 °C annealed film exhibits the shortest phase-coherence length decreasing from 123 nm at 2.5 K to 47 nm at 10 K. For all three films, $L_e$ is in the range of 8-9 nm, corresponding to $B_{tr}$ ~ 4-5 T, well above the 2T field range probed in our magnetoconductance measurements. We further fit the temperature dependence of $\mathcal{L}_{\varphi}$ for all the three films to the power law equation $\mathcal{L}_{\varphi} \propto T^{-p}$, consistent with the Nyquist theory of expectations for dephasing in a diffusive two-dimensional system [62,63] . From the fitting, we obtain the value of p = 0.7. A value of p = 0.5 is expected for electron-electron scattering while p = 1 corresponds to electron-phonon interaction [22,64] . Our intermediate value of p = 0.7 suggests that both mechanisms contribute to the dephasing, with electron-electron scattering playing the major role.

SnTe hosts four Dirac cones on each high-symmetry surface [5,22] . Assuming that both the top and bottom surfaces of the film contribute independently, the theoretical maximum for 2α could reach 8 [65,66]. However, experimental values are typically much lower because the intervalley and intravalley scattering can couple the channels and reduce the effective degeneracy. As shown in Fig. 4(a), the 2α values for the 150 °C and 200 °C annealed samples at 2.5 K are 2.84 and 2.74 respectively. In contrast, the 100 °C annealed sample exhibits a significantly higher 2α of approximately 3.5 at the same temperature. The relatively higher values are comparable to those reported previously for SnTe thin films [22,65]. The relatively higher value of 2α in the 100 °C film suggests that lower disorder allows a larger number of coherent transport channels to remain effectively independent, consistent with longer phase coherence length observed in this sample. The validity of the HLN model in the strong SOC limit is further supported by the characteristic lengths extracted from the fits. We first convert the sheet resistance to a magnetoconductance, $\Delta G(B) = \frac{1}{R_S(B)} - \frac{1}{R_S(0)}$. Each ΔG (B) curve is then fitted to Eq. (1) with α, $B_{\varphi}$, and β as free parameters. The dephasing field $B_{\phi} = \frac{\hbar}{4e\mathcal{L}_{\phi}^2}$ is obtained directly from the fit, yielding $\mathcal{L}_{\phi}$ at each temperature. The strong spin orbit interaction requires that the phase coherence length $\mathcal{L}_{\phi}$ and $L_{SO}$ satisfy the condition $\mathcal{L}_{\phi} \gg L_{SO}$ across the entire measurement range [10,67] . To extract $L_{so}$, we separate the quantum contribution $\beta_q$ from the fitted β. We independently determine $\beta_c = \mu^2 B^2$ from the Hall mobility and carrier density, and obtain $\beta_q = \beta - \beta_c$. Using $\beta_q$ together with $B_{\varphi}$ and $B_e$ in Eq. (2), we solve for $B_{SO}$, and obtain the spin-orbit scattering length as $L_{so}$ = 16.54 nm. Figure 4(b) compares the two coherence lengths for the 200 °C annealed sample as a function of temperature. While $\mathcal{L}_{\phi}$exhibits a characteristic power-law decrease with temperature – typical of dephasing dominated by inelastic scattering – the spin-orbit scattering length remains nearly temperature independent. This further supports the fact that the transport in the strong spin-orbit coupling regime.

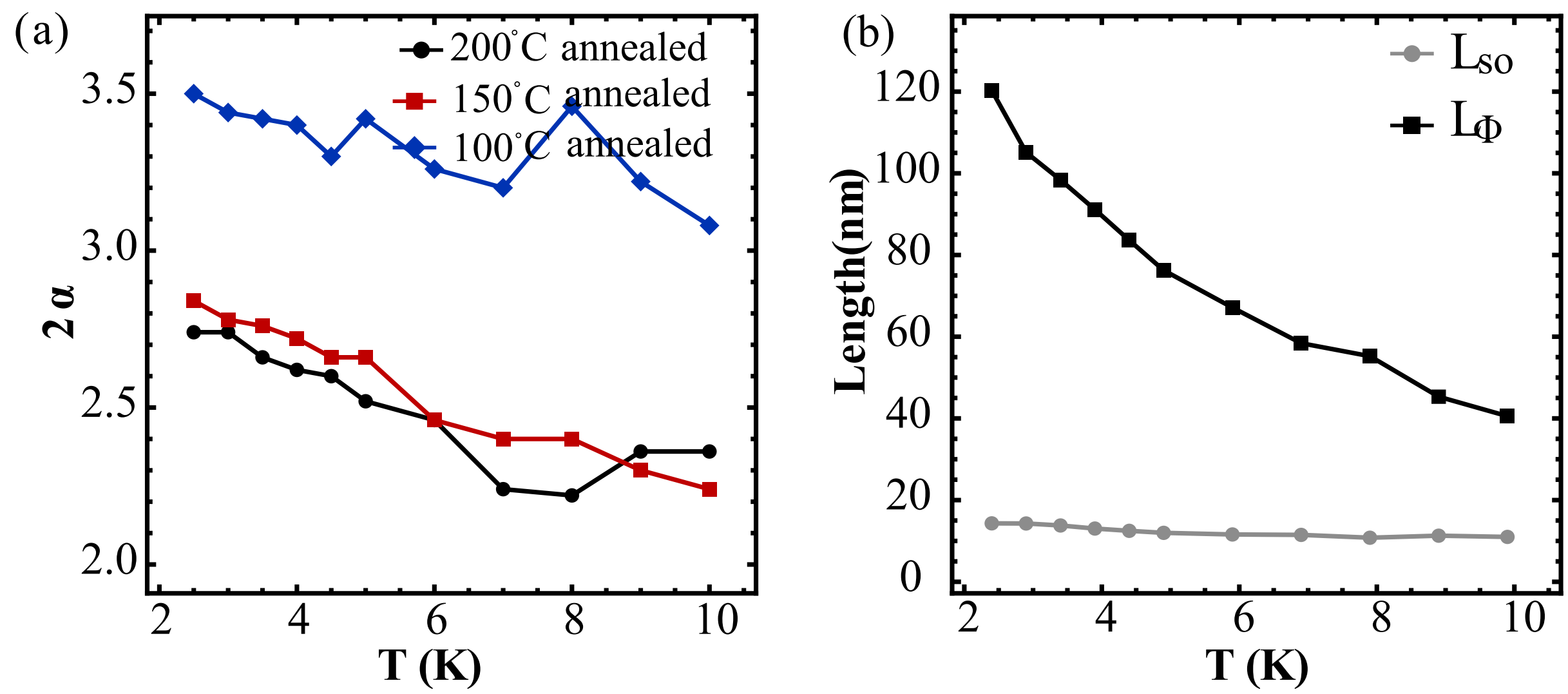


FIG.4. Temperature dependence of the WAL parameters in SnTe thin films. (a) Temperature dependence of the weak antilocalization prefactor 2α representing the effective number of independent transport channels, for the films annealed at 200, 150 and 100 °C. The 100 °C annealed film exhibits the largest 2α over the measured temperature range, indicating weaker coupling between the Dirac-valley transport channels. (b) Temperature dependence of the phase coherence length $\mathcal{L}_\phi$ and the spin-orbit scattering length $L_{so}$ for the 200 °C annealed film. $\mathcal{L}_\phi$ decreases with increasing temperature, while $L_{SO}$ remains comparatively weakly temperature dependent. The persistent condition of $\mathcal{L}_\phi >> L_{so}$ is consistent with the strong spin-orbit-coupling regime assumed in the HLN analysis.

Having discussed the magneto-transport at T ≤ 10 K, it is pertinent to see the behavior of MR in the temperature range where quantum effects are less relevant. The percentage of the magnetoresistance is defined as

$$MR\ \% \ = \frac{\rho\ (H) - \rho\ (0)}{\rho(0)} X\ 100\% \quad (3)$$

where ρ(H) and ρ (0) are the resistivity measured in the presence and absence of magnetic field.

Figure 5(a), 5(e), and 5(i) show the magnetoresistance of the SnTe thin films. It is noted that in all samples the MR increases with the strength of the magnetic field and decreases with increasing temperature.

To probe how the dominant scattering processes evolve with temperature and field, we analyze the MR data using Kohler's rule. The original Kohler's rule states that MR = f(H/$\rho_o$), where $\rho_o$ is the resistivity at zero magnetic field (H). Within a semiclassical picture of transport, the field – induced change in resistivity depends on the product Hτ, where τ denotes the mean time between successive scattering events for the conduction electrons. For a single-band system, $\rho_o = \frac{m^*}{ne^2\tau}$. When the carrier density n and effective mass m* are temperature independent, $\rho \propto \frac{1}{\tau}$ , and the ordinary Kohler scaling can be expressed as H/$\rho_o$. As a result of this, the MR curves measured at different temperatures should collapse onto a single universal curve when plotted as a function of H/$\rho_o$. As seen in Figs. 5(b), 5(f), and 5(j), this scaling law of MR does not collapse the data at all temperatures into a single universal curve, showing a clear violation. Such deviations can arise

when the assumptions underlying conventional Kohler's rule are not satisfied, for example due to temperature-dependent carrier density, multiple carrier contributions, or more than one relevant scattering time [68,69] .

We use the extended Kohler's rule as proposed by Xu et al. [69]where an additional factor of $1/n_T$ is used to account for the temperature-dependent carrier density. Using the equation $\rho_o = m^*/ne^2\tau$, where τ is the relaxation time, we can write $H\tau = (m^*/e^2)\ H/\ n\ \rho_o$. If the effective mass m* is temperature-independent then the extended Kohler's rule is given as

$$MR = f\left(\frac{H}{n_T\,\rho_O}\right) \quad (4)$$

We emphasize that $n_T$ is introduced in the extended Kohler framework as a thermal scaling factor associated with the temperature dependence of the carrier density. As shown by Xu et al., $n_T$ can reflect the temperature dependence of the carrier density when the densities or mobilities of the contributing bands have the same or similar temperature dependence; otherwise $n_T$ contains contributions from the temperature dependences of both the carrier densities and mobilities [69] In the present analysis, however, we also treat $n_T$ empirically and compare it with the independently determined Hall carrier density to examine whether the temperature dependence of the carrier density alone can account for the observed MR scaling. The Hall resistivity of the films is linear in magnetic field over the entire temperature range (2.5 to 300 K), allowing the Hall carrier density $n_{Hall}$ (T) to be obtained within a single-carrier approximation. For all three films, the normalized Hall-derived carrier density factor $n_{\,Hall} = n\,(T)/n\,(250\text{ K})$ remains close to unity. We initially use this quantity as a physically motivated estimate of $n_T$. However, the resulting plots in Figs. 5 (c), 5 (g), and 5 (k) show only a modest improvement over conventional Kohler scaling, indicating that the temperature dependence of the Hall-derived carrier density alone is insufficient to account for the observed MR scaling.

Having found that the ordinary Kohler scaling fails (Figs. 5(b), 5(f), and 5(j)) we asked whether the failure could be repaired by the simplest possible fix: sliding the MR curve for each temperature sideways by a single number $n_T$, while keeping one shared power law shape for all temperatures. We assumed that at every temperature the MR(H) curve follows the same power law, MR = A (T). $H^m$ with one shared m across all temperatures – only the prefactor A(T) is allowed to change with T. Under this assumption the value of $n_T$ needed to bring each curve's amplitude A(T) into the line with the 250 K reference curve is fixed by the algebra alone: $n_T = \frac{\rho_O(250\,K)}{\rho_O(T)}\left(\frac{A_{250}}{A_T}\right)^{1/m}$. We computed this $n_T$ for each temperature and used it to rescale the field axis, plotting MR against the $H/n_T\,\rho_o$ (dashed lines in Figs. S4(a)-S4(c) of the Supplemental Material [42], with m=0.9, 2.1, and 0.8 for the 200, 150 and 100 °C films, respectively). This approach also fails to collapse the curves because $n_T$ only rescales the field axis and cannot account for different slopes of the MR curves. No single m can therefore simultaneously match both the vertical offset and the shape of every curve.

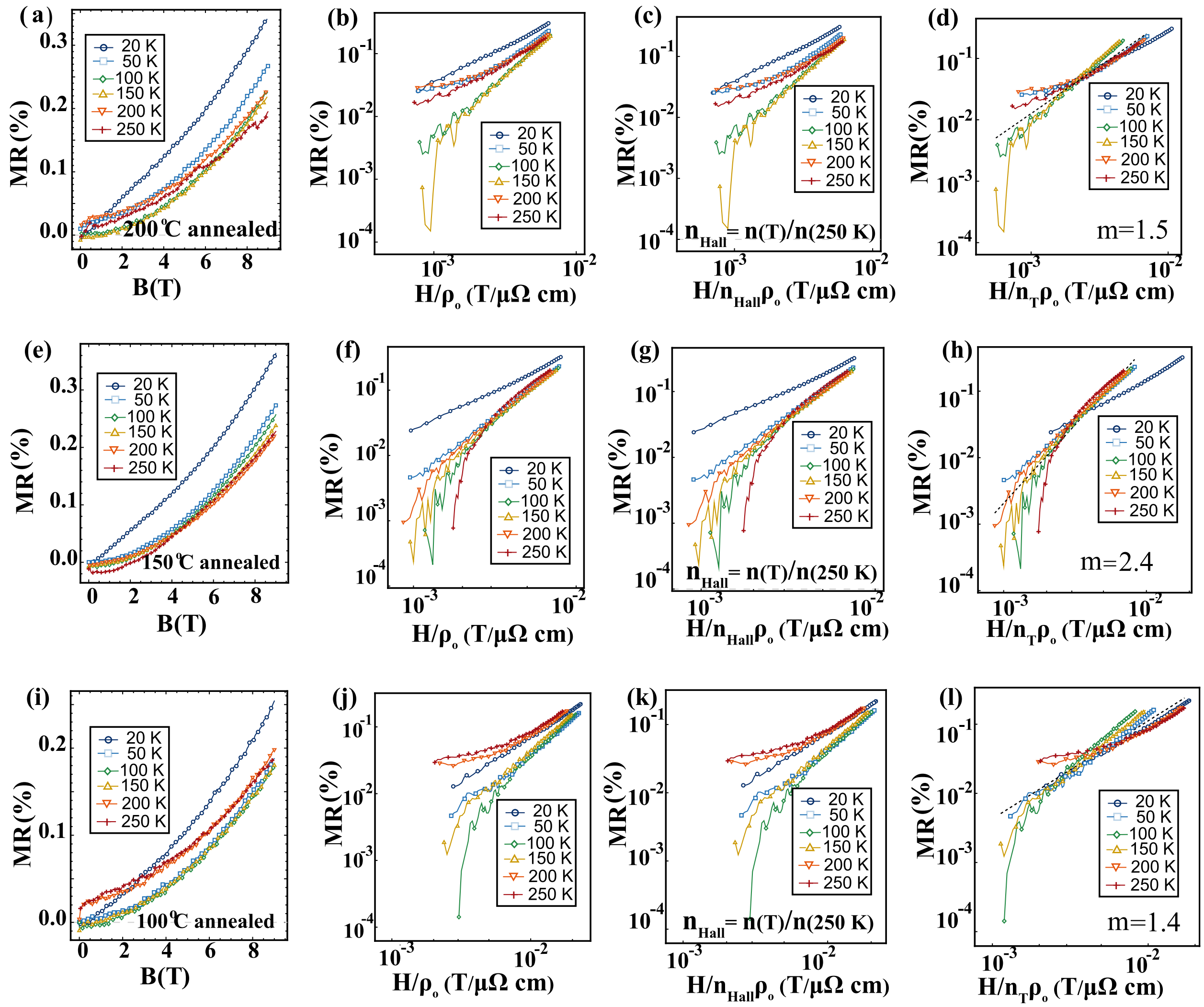

FIG.5. Magnetoresistance and Kohler scaling analysis of SnTe thin films annealed at (a)-(d) 200°C, (e)-(h) 150° C, and (i)-(l) 100° C. (a), (e), and (i) show the field dependent MR at temperatures between 20 and 250 K for the three films. (b), (f) and (j) present conventional Kohler plots of MR versus $H/\rho_o$. (c), (g) and (k) show the corresponding modified Kohler plots using the Hall carrier density, $n_{Hall}(T)/n_{Hall}(250\ K)$, as the scaling factor. (d), (h) and (l) display extended Kohler scaling using $n_T$ obtained by collapsing each curve onto the 250 K reference curve, overlaid with the best fit power law MR $\propto [H/(n_T \rho_o)]^m$ (solid lines), where m = 1.54 (200° C), 2.36 (150°C), and 1.42 (100° C). The temperature-dependent MR curves exhibit varying degrees of scaling collapse, with the 20 K data deviating from the scaling curve in all films.

We therefore determine $n_T$ empirically, by collapsing each MR curve onto the 250 K data in the window 0-9 T with no assumed exponent or equating it with $n_{Hall}$. The collapsed plots are Figs. 5(d), 5(h), and 5(l). The values of $n_T$ to achieve scaling in all three samples are listed in Table I. $n_T$ and Hall ratio $n_{Hall}$ are compared in Figs. S4(d)-S4(f) of the Supplemental Material [42]. It may be noted that the data at 20 K remain off the main scaling trajectories, indicating the onset of a different regime of carrier scattering. Since we previously established WAL induced MR in the temperature range of 2.5 to 10 K, the deviation from scaling at 20 K may indicate a crossover toward the low-temperature regime in which quantum interference or additional temperature-dependent scattering processes become relevant. However, the Kohler analysis alone does not allow the deviation at 20 K to be uniquely attributed to WAL.

Table I. Values of the empirical thermal scaling factor $n_T$ obtained by collapsing the MR(H) curves at each temperature onto the 250 K reference curve [Figs. 5(d), 5(h), and 5(l)].

| T (K) | 200 °C | 150 °C | 100 °C |
|---|---|---|---|
| 20 | 0.6 | 0.45 | 1.25 |
| 50 | 0.9 | 0.95 | 2.05 |
| 100 | 1.4 | 1 | 2.55 |
| 150 | 1.5 | 1 | 2.1 |
| 200 | 0.95 | 1 | 1.12 |
| 250 | 1 | 1 | 1 |

# IV. CONCLUSION

We have successfully grown highly textured SnTe films on c-plane sapphire ($Al_2O_3$) by pulsed laser deposition and investigated the effect of post-growth annealing on their structural and magneto-transport properties. XRD measurements confirmed the formation of the cubic SnTe phase and showed improved crystallinity and a stronger (111) preferred orientation with increasing annealing temperature, while Raman and XPS measurements further supported the formation of SnTe, while EDS shows that the 100 °C film is Te-rich, and the films annealed at higher temperatures approach 1:1 Sn:Te stoichiometry. All films are p-type, with carrier concentrations on the order of $10^{20}$ $cm^{-3}$; both carrier concentration and Hall mobility are highest in the 100 °C film. The 200 °C film is the most resistive and exhibits a broad, weakly field-dependent resistivity minimum near 125 K, described within a metallic granular-transport framework. At low temperatures, weak antilocalization (WAL) is observed in all the films and is well described by the HLN model. The Te-rich 100 °C film exhibits the longest phase coherence length (134 nm) and the largest number of coherent transport channels, (2α~3.5 at 2.5 K). The phase-coherence length follows $\mathcal{L}_\phi \propto T^{-0.7}$, indicating contributions from both electron-electron and electron-phonon scattering to dephasing. At higher temperatures, an improved collapse of the magnetoresistance is obtained using the extended Kohler approach with an empirical thermal scaling factor. The 100 °C film shows the largest variation in this factor, while deviations in the 200 °C film are more pronounced at 100 and 150 K. The 150 °C film shows the most consistent scaling, with the scaling factor remaining close to unity over most of the measured temperature range. The deviation at 20 K suggests a crossover between the low temperature quantum-interference regime and the high-temperature transport behavior. Taken together, the 150 °C film combines clear WAL, the most consistent magnetoresistance scaling, and a composition close to 1:1 Sn:Te. Overall, these results demonstrate that post-growth annealing modifies the crystallinity and composition of SnTe films while simultaneously influencing dephasing, phase coherence, and classical magnetotransport, revealing a close interplay between structural refinement and quantum transport.

## ACKNOWLEDGMENTS

This work was funded by the Department of Defense through Grant Number W911NF2120213 to the DOD Center of Excellence for Advanced Electro-Photonics with 2D Materials at Morgan State University. We thank Dr. Shiva Pokhrel for assistance with the XRD measurements at Towson University.